\documentclass[a4paper,11pt]{article}
\usepackage{jcappub} 
\usepackage{amsmath}

\title{\boldmath Dynamical analysis of cosmological models in non-minimal scalar non-metricity gravity}

\author{Bikash Chandra Roy, Bikash Chandra Paul}
\affiliation{Department of Physics, University of North Bengal, Raja Rammohunpur, Darjeeling, Siliguri-734013, West Bengal, India}

\emailAdd{bcroy.bcr25@gmail.com, bcpaul@nbu.ac.in}

\abstract{We investigate the evolution of a spatially flat Friedmann-Robertson-Walker (FRW) universe in the framework of scalar non-metricity theory of gravity. In the model, we consider  dark matter (DM) and dark energy (DE) described by the scalar field. In the paper, we examine the asymptotic behavior of the evolution of the observed universe. We probe the universe with and without an interaction between DM and the effective energy density including scalar field in the non-metricity theory that describes the DE. The critical points of the autonomous system and their stability are determined to understand the behaviour of the universe. The cosmological models accommodate the late accelerating phase of the universe for a given set of parameters, a matter-dominated saddle point is also found to exist which is flowed by a decelerating phase dominated by stiff fluid like DE. We obtain a new class of cosmological model accommodating early inflation, matter-dominated era, and a future decelerating phase followed by late acceleration of the universe.}

\begin{document}
\maketitle
\flushbottom

\section{Introduction}
\label{intro}
The astronomical observations in the last few decades, namely the type Ia supernovae \cite{riess_observational_1998,1999ApJ...517..565P,riess2004type}, Cosmic Microwave Background Radiation (CMBR) \cite{Spergel_2003,Spergel_2007,Komatsu_2009,Komatsu_2011,Planck_Collaboration_2020}, Baryonic Acoustic Oscillations (BAO) \cite{Eisenstein_2005} and large-scale structure (LSS) \cite{PhysRevD.69.103501,PhysRevD.74.123507,PhysRevD.71.103515} have confirmed the most astounding observational result that the universe is not only expanding but it is passing through an era of accelerated expansion which is termed as late acceleration. The universe has undergone a decelerating expansion, being dominated first by radiation and then by matter. It is also known in modern cosmology that the present universe emerged from an inflationary phase of expansion at very early times \cite{Guth1981}. The inflationary phase was required to get rid of the issues in Big Bang cosmology when probed in the early era. The basic problems that are resolved are flatness, horizon problems, it also provides an explanation of the observed large-scale structure (LSS) \cite{Abazajian_2015}. It is predicted from the present astronomical observation that the universe is expanding can be understood with fliud having unique characteristic of negative pressure but the fields that are required to describe the early universe are different, which is called DE. Initially, the late-time acceleration of the universe is developed theoretically making use of the cosmological constant $\Lambda$ in Einstein’s field equations, that yields a cosmology known as $\Lambda$ cold dark matter ($\Lambda$CDM) model provides an excellent fit of the observations. The drawback of model is that it fails to explain two major issues: the cosmological constant problem, which refers to the enormous discrepancy between the theoretical and observed values of $\Lambda$ and the coincidence problem, which arises from the fact that the present-day energy densities of DE ($\rho_{\Lambda}$) and matter ($\rho_{m}$) are of the same order of magnitude despite evolving very differently over time. Consequently, a number of dynamical DE models, namely quintessence \cite{PhysRevD.37.3406,chiba_quintessence_1999}, tachyon \cite{Padmanabhan_2002,Bagla_2003,srivastava2005tachyondarkenergysource}, k-essence \cite{PhysRevD.63.103510}, phantom \cite{Caldwell_2002,Caldwell_2003}, Chaplygin gas \cite{Kamenshchik_2001} etc., are introduced within GR to address the present issues in cosmology.

The late acceleration of the universe is also explored  modifying the gravity sector of the Einstein field equation  adding extra terms in the Einstein-Hilbert action. In the literature, modified theories of gravity, namely, $f(R)$ gravity in which the Lagrangian density is an arbitrary function of $R$ which are important at the late epoch \cite{Sotiriou_2010,De_Felice_2010}; scalar-tensor theory where a scalar field is nonminimally coupled to curvature \cite{Horndeski1974}; $f(R,\mathcal{T})$ gravity in which the Lagrangian density is written with a nonminimal coupling between geometry and matter \cite{PhysRevD.84.024020}; Einstein-Gauss-Bonnet gravity and generalizations with $f(\mathcal{G})$ and $f(R,\mathcal{G})$ \cite{NOJIRI20051,De_Felice_2009}; Horava-Lifshitz gravity \cite{wang_horava_2017}; brane world gravity \cite{DVALI2000208,Maartens_2010} etc. are considered to study cosmological issues to obtain realistic cosmological models \cite{nojiri_modified_2006,Nojiri_2011,refId0,RUDRA2021115428,SURANJOYSINGH2021101542,Fortunato2024,PhysRevD.76.044027,ODINTSOV2019935,Chandapaul_2023,Chakrabarti_2012,MAITY2025103026,Zonunmawia_2018}. 

It is known that General Relativity (GR) and its extensions results from the geometrical interpretation of gravity, where the gravitation is describe by the curvature as
\begin{equation}
    R^{\alpha}_{\ \beta\mu\nu} \equiv \partial_{\mu}\Gamma_{\ \nu\beta}^{\alpha} - \partial_{\nu}\Gamma_{\ \mu\beta}^{\alpha} + \Gamma_{\ \mu\lambda}^{\alpha}\Gamma_{\ \nu\beta}^{\lambda} - \Gamma_{\ \nu\lambda}^{\alpha}\Gamma_{\ \mu\beta}^{\lambda},
\end{equation}
here the affine connection $\Gamma^{\alpha}_{\ \mu\nu} = \{^{\alpha}_{\mu\nu}\}$, is the metric compatible and torsionless Levi-Civita connection defined in terms of metric $g_{\mu\nu}$ as
\begin{equation}
    \{^{\alpha}_{\mu\nu}\} = \frac{1}{2}g^{\alpha\lambda}\left(\partial g_{\nu\lambda,\mu} + \partial g_{\mu\lambda,\nu}- \partial g_{\mu\nu,\lambda}\right).
\end{equation}
However, gravity can be reformulated as the teleparallel equivalent of general relativity (henceforth, TEGR) based on geometrical object torsion, generated by tetrad fields which satisfactorily described the gravitational effect \cite{pellegrini1963tetrad,RevModPhys.48.393,PhysRevD.19.3524}. In this formulation, the Weitzenböck connection is used, which yields the vanishing curvature $R^{\alpha}_{\beta\mu\nu} = 0$ and the non-metricity term $Q_{\alpha\mu\nu} = \nabla_{\alpha}g_{\mu\nu} = 0$. Although TEGR is considered completely equivalent to GR, it has distinct characteristics that make it an attractive framework for further development, one such mechanism is its gauge-theoretic structure and the ability to separate inertial and gravitational effects \cite{PhysRevD.99.064006}. It is realized that the new formulation of gravity, still have issues that are encountered in cosmology  using GR. Therefore, an extension of TEGR making modifications with the Lagrangian density containing an arbitrary function of $T$, namely, $f(T)$ gravity  \cite{bengochea_dark_2009,Linder_2010,Maluf_2013}. An important advantage of $f(T)$ gravity is that its field equations are of the second-order differential equation, unlike fourth-order differential equations in $f(R)$ gravity. Recently, $f(T)$ gravity is taken to investigate both the early inflationary phase \cite{PhysRevD.75.084031,Awad_2018} and the late accelerating  universe \cite{Wu_2010,Yang_2011,Bamba_2011,Capozziello_2011,Myrzakulov_2011,PhysRevD.85.104036,capozziello2017constraining,Chakrabarti_2017} (Review of various torsional constructions, cosmological and astrophysical applications can be found in \cite{Cai_2016}). Another approach has been made with this formulation making use of the coupling scalar fields to torsion and a non-minimal coupling between geometry and matter. This leads to the emergence of distinct classes of gravitational theories, such as scalar-torsion theory of gravity \cite{Geng_2011,PhysRevD.89.084025}, $f(T,\mathcal{T)}$ gravity \cite{Harko_2014}. These theories are found to  exhibit some interesting cosmological characteristics \cite{PhysRevD.111.043506,DUCHANIYA2024101402,Momeni_2014,Farrugia_2016,Junior_2016}.

It is also realized that another representation of GR is possible adopting a connection with vanishing curvature and torsion, which is called symmetric teleparallel formulation of general relativity (henceforth, STGR) \cite{nester1999symmetric}, (see also \cite{adak2004solution,ADAK_2006,ADAK_2013}). In STGR, the basic geometric variable describing the properties of the gravitational interaction  represented by the non-metricity, which is defined by $Q_{\alpha\mu\nu} = \nabla_{\alpha}g_{\mu\nu}$ with the variation of the length of a vector in parallel transport. Different geometrical and physical aspects of STGR have been investigated in the past two decades in a number of studies \cite{adak2004solution,adak2006stg}. By an approprite choice of gauge namely, the coincidence gauge \cite{Jim_nez_2018}, the symmetric teleparallel formulation is found to reduce to the field equations that obtained from GR. The theory obtained above is called the symmetric teleparallel equivalent of general relativity (STEGR). Further, the STEGR was extended into the $f(Q)$ gravity theory, where $Q$ is the non-metricity scalar \cite{Jim_nez_2018}. A volume of work resulted in $f(Q)$ gravity, such as test of  energy conditions \cite{mandal2020energy}, bouncing cosmologies \cite{bajardi2020bouncing}, cosmography \cite{PhysRevD.102.124029}, quantum cosmology \cite{Dimakis_2021}, covariant formulation \cite{Zhao_2022}. In the literature \cite{lu2019_1906.08920,lazkoz2019observational,PhysRevD.101.103507,frusciante2021signatures,anagnostopoulos2021first,Atayde_2021,PhysRevD.103.063505,khyllep2021cosmological,koussour2022late,Arora_2022,Solanki_2022,Narawade_2023,Narawade_2023_a3,De_2023,rana_2024a} $f(Q)$ gravity is used to investigate a number of cosmological issues and their observational constraints. The cosmological perturbation in $f(Q)$ gravity is studied \cite{PhysRevD.101.103507} to explore  the re-scaling of the Newton's Gravitational constant in tensor perturbations and the absence of vector contributions. The scalar sector in the perturbation introduced two additional propagating modes, predicting at least two extra degrees of freedom in $f(Q)$. The growth index of matter perturbations has been studied in the context of $f(Q)$ gravity \cite{khyllep2021cosmological}. The modified $f(Q)$ gravity is found to support the constraints imposed by Big Bang Nucleosynthesis (BBN) \cite{anagnostopoulos2023new} and a detail review of $f(Q)$ gravity is given in Ref. \cite{HEISENBERG20241}. To address the recent acceleation of the universe there is a spurt in research activities in $f(Q)$ gravity which is an alternative framework for late universe understanding. It is shown \cite{gomes2024pathological,heisenberg2023perturbations} that a nonlinear extension of $f(Q) \neq Q$ is non-viable for cosmological applications due to strong coupling or ghost instability. To avoid the ghost in $f(Q)$ gravity it  requires not more than two propagating gravitational degrees of freedom and avoidance of inclusion of the higher-order derivative terms in the field equations thus admitting  only  linear form of $f(Q)$ theory. Furthermore, non-metric gravity was extended by incorporating the coupling between geometry and matter. In this framework, the gravity is formulated in $f(Q, \mathcal{L}_{m})$ \cite{PhysRevD.98.084043} and $f(Q, \mathcal{T})$ \cite{xu2019f,xu2020weyl}, where the nonmetricity scalar $Q$ non-minimally coupled to the matter Lagrangian $\mathcal{L}_{m}$ and the trace of the matter-energy-momentum tensor $\mathcal{T}$, respectively. Myrzakulov {\it et al.} obtained analytical solutions and fit the late-time cosmology with observational data in the framework of $f(Q, \mathcal{L}_{m})$ gravity \cite{MYRZAKULOV2024101614}. The observational constraints of the the model parameters are determined in the literature \cite{Myrzakulov_2025,hazarika_2025}. In $f(Q, \mathcal{T})$ gravity theories, extensive studies have been carried out in the last few years, including models of the universe in various cosmological scenarios \cite{xu2019f,xu2020weyl,arora2020f,Koussour_2022,Das_2024}, energy conditions \cite{arora2020energy,ARORA2021100790}, cosmological perturbations \cite{N_jera_2022}, cosmological inflation \cite{Shiravand_2022}, holographic DE \cite{doi:10.1142/S0219887823500846}. In addition to this, there is another extension of STEGR gravity by incorporating a scalar field that is non-minimally coupled to the non-metricity scalar $Q$, called scalar non-metricity theory \cite{J_rv_2018}. The scalar-nonmetricity theory do not contain higher-order derivatives of $Q$, consequently  the linear term in $Q$ considered here which is cosmologically viable and ghost-free when $\mathcal{A}(\phi)$ and its derivatives of all orders are continuous.

In the present paper, we investigate cosmological model based on scalar non-metricity theory of gravity for a flat FRW universe using a dynamical system analysis for the different phases of the universe. We adopt the autonomous technique to analyze the field equations as these are highly non-linear. The viability and stability of the cosmological models are tested by analyzing the local asymptotic behavior of critical points (C.P.) or fixed points connecting them to the major cosmological epochs of the universe. For example, the radiation and matter-dominated periods correlate to saddle points, but late-time (the DE sector) dominance normally corresponds to a stable point. Such an autonomous dynamical system approach has been applied to obtain cosmological models with different dynamical DE  \cite{PhysRevD.57.4686,doi:10.1142/S021827180600942X},  and  the modified theories of gravity \cite{Amendola_2007,Koivisto_2010,Uddin_2009}.

The paper is organized as follows: In section \ref{FE}, we present the theoretical framework of the scalar non-metricity theory of gravity for a flat FRW Universe. In section \ref{sec:dyn}, we construct the dynamical system corresponding to the scalar non-metricity theory and investigate the conditions under which it can be autonomous; as it is proved, the cases for which the dynamical system is rendered autonomous correspond to distinct cosmological scenarios. In section \ref{sec:4}, we find the corresponding critical points for the autonomous system in the cases of (i) without interaction and (ii) with interaction and discuss their stability and bifurcations and understanding the flow of the system in the phase diagram. Additionally, we also find out the numerical results from the integration of the dynamical system. Finally, in section \ref{sec:conclusions}, we provide an extensive summary of the paper.

\section{The field equations}
\label{FE}
The action functional for scalar non-metricity theory of gravity can be written as \cite{J_rv_2018}
\[\mathcal{S} = \frac{1}{2}\int d^{4}x\sqrt{-g} \Big[\mathcal{A}(\phi)Q - \mathcal{B}(\phi)g^{\alpha\beta}\partial_{\alpha}\phi\partial_{\beta}\phi - 2V(\phi)\]
\begin{equation}\label{FE:sn1}
     \hspace{1.5 cm}+ 2\lambda_{\mu}^{\;\beta\alpha\gamma} R^{\mu}_{\;\beta\alpha\gamma} + 2\lambda_{\mu}^{\;\alpha\beta}T^{\mu}_{\;\alpha\beta}\Big] + \mathcal{S}_{m},
\end{equation}
where $g$ is the determinant of the metric tensor $g_{\mu\nu}$, $\mathcal{A}(\phi)$ is a generic function of the scalar field $\phi$ non-minimally coupled to the non-metricity scalar $Q$, defined as
\begin{equation}
    Q = - \frac{1}{4}Q_{\alpha\beta\gamma}Q^{\alpha\beta\gamma} + \frac{1}{2}Q_{\alpha\beta\gamma}Q^{\gamma\beta\alpha} + \frac{1}{4}Q_{\alpha}Q^{\alpha} - \frac{1}{2}Q_{\alpha}\tilde{Q}^{\alpha},
\end{equation}
with non-metricity tensor
\begin{equation}
    Q_{\alpha\mu\nu} = \nabla_{\alpha}g_{\mu\nu}
\end{equation}
and $Q_{\alpha} \equiv Q_{\alpha}^{\;\;\mu}{}_{\mu}$, $\tilde{Q}^{\alpha} \equiv Q_{\mu}^{\;\;\mu\alpha}$ are the trace of the non-metricity tensor. One can obtains explicitly the form of $Q$ as
\begin{equation}
    Q = - Q_{\alpha\mu\nu}P^{\alpha\mu\nu}, 
\end{equation}
where $P^{\alpha}{}_{\mu\nu}$ is the non-metricity conjugate tensor (or superpotential) and defined as
\begin{equation}
    P^{\alpha}{}_{\mu\nu} \equiv -\frac{1}{2}L^{\alpha}{}_{\mu\nu} + \frac{1}{4}(Q^{\alpha} - \tilde{Q}^{\alpha})g_{\mu\nu} - \frac{1}{4}\delta^{\alpha}{}_{(\mu}Q_{\nu)}.
\end{equation}
In the above equation $L^{\alpha}_{\mu\nu}$ is the  disformation tensor which measures the separation of the Levi-Civita connection the symmetric part of the full connection with it and can be expressed as
\begin{equation}\label{FE:3}
    L^{\alpha}{}_{\mu\nu} \equiv \frac{1}{2}\left(Q^{\alpha}{}_{\mu\nu} - Q_{\mu\;\;\nu}^{\;\alpha} - Q_{\nu\;\;\mu}^{\;\alpha}\right).
\end{equation}
The function $\mathcal{B}(\phi)$ is the generic kinetic coupling of the scalar field (for canonical scalar field $\mathcal{B}(\phi) = 1$) and $V(\phi)$ is the self-interaction potential of the scalar field. In this theory, $\mathcal{S}_{m} = \mathcal{S}_{m}(g_{\mu\nu},\chi)$ denotes the action of matter fields $\chi$, the quantities $R^{\mu}_{\;\;\beta\alpha\gamma}$, $T^{\mu}_{\;\;\alpha\beta}$ are the Riemann curvature and torsion tensor, respectively. The $\lambda_{\mu}^{\;\;\beta\alpha\gamma}$ and $\lambda_{\mu}^{\;\;\alpha\beta}$ are two Lagrange multipliers that are assumed to be anti-symmetrized, i.e., $\lambda_{\mu}^{\;\;\beta\alpha\gamma} = \lambda_{\mu}^{\;\;\beta[\alpha\gamma]}$ and $\lambda_{\mu}^{\;\;\alpha\beta} = \lambda_{\mu}^{\;\;[\alpha\beta]}$ and are introduced into the action to enforce constraints on curvature and torsion. The antisymmetrized Lagrange multipliers ensure curvature $R_{\beta\alpha\gamma}^{\mu} = 0$ and torsion $T_{\alpha\beta}^{\mu} = 0$, as expected in the symmetric teleparallel framework \cite{J_rv_2018}. In the case when $\mathcal{A} = 1$ and $\mathcal{B} = \mathcal{V} = 0$ the theory reduces to STEGR. 

Varying the action (\ref{FE:sn1}) with respect to the metric $g_{\mu\nu}$ and in the symmetric teleparallel formulation, yields
\[\mathcal{T}_{\mu\nu} = \frac{2}{\sqrt{- g}}\nabla_{\alpha}\Big(\sqrt{- g}\mathcal{A}P^{\alpha}_{\;\;\mu\nu}\Big) - \frac{1}{2}g_{\mu\nu}\mathcal{A}Q + \mathcal{A}\Big(P_{\mu\alpha\beta}Q_{\nu}^{\;\;\alpha\beta} - 2Q_{\alpha\beta\mu}P^{\alpha\beta}_{\;\;\;\nu}\Big)\]
\begin{equation}\label{Eq:sng3}
\hspace{1cm}+ \frac{1}{2}g_{\mu\nu}\Big(\mathcal{B}g^{\alpha\beta}\partial_{\alpha}\phi\partial_{\beta}\phi + 2V\Big) - \mathcal{B}\partial_{\mu}\phi\partial_{\nu}\phi,
\end{equation}
where $\mathcal{T}_{\mu\nu}$ is the matter energy-momentum tensor defined as
\begin{equation}
    \mathcal{T}_{\mu\nu} = -\frac{2}{\sqrt{- g}}\frac{\delta\sqrt{- g}\mathcal{L}_{m}}{\delta g^{\mu\nu}},
\end{equation}
where $\mathcal{L}_{m}$ is the matter Lagrangian. Again, by variation the action (\ref{FE:sn1}) with respect to the scalar field $\phi$, one can obtain the scalar field equation as
\begin{equation}\label{}
    2\mathcal{B}\nabla_{\alpha}\nabla^{\alpha}\phi + \mathcal{B}'g^{\alpha\beta}\partial_{\alpha}\phi\partial^{\beta}\phi + \mathcal{A}'Q - 2V' = 0,
\end{equation}
where the primes denotes the derivative with respect to the scalar field $\phi$. Like scalar-tensor and scalar-torsion theory, the equation obtains a term with the geometric invariant to which the scalar field is non-minimally coupled. Note that for minimal coupling, $\mathcal{A}(\phi) = 1$ and canonical scalar field, $\mathcal{B}(\phi) = 1$, the above equation reduce to standard Klein-Gordon equation.

Now, consider the spatially flat Friedmann-Robertson-Walker (FRW) spacetime, whose metric is of the form
\begin{equation}\label{Eq:RW1}
    ds^{2} = - dt^{2} + a(t)^{2}\delta_{ij}dx^{i}dx^{j},
\end{equation}
where $t$ is the cosmic time, $a(t)$ is the scale factor. When the connection is zero (the coincident gauge), the non-metricity scalar is $Q = 6H^{2}$ with $H \equiv \frac{\dot{a}}{a}$ is the Hubble parameter and dot stands for a derivative with respect to $t$. Also, we assume that the matter is a perfect fluid whose energy-momentum tensor $\mathcal{T}_{\mu\nu}$ is given by
\begin{equation}
    \mathcal{T}_{\mu\nu} = (\rho + p)u_{\mu}u_{\nu} + pg_{\mu\nu},
\end{equation}
where $u_{\mu}$ is the four-velocity satisfying the normalization condition $u_{\mu}u^{\mu} = -1$, $\rho$ and $p$ are the energy density and pressure of a perfect fluid respectively.

By imposing $\mathcal{A}(\phi) = 1 + \mathcal{F}(\phi)$ and considering the case that the Universe is filled with DM, the modified Friedmann equations are
\begin{equation}\label{Eq:Friedmann_1}
    3H^{2} = \rho_{m} + \frac{1}{2}\dot{\phi}^{2} + V - 3\mathcal{F}H^{2},
\end{equation}
\begin{equation}\label{Eq:Friedmann_2}
    (1 + \mathcal{F})(2\dot{H} + 3H^{2}) = - 2\mathcal{F}'H\dot{\phi} - \frac{1}{2}\dot{\phi}^{2} + V - p_{m},
\end{equation}
where $\rho_{m}$ and $p_{m}$ are the energy density and pressure of DM (usually $p_{m} = 0$). We rewrite equations, (\ref{Eq:Friedmann_1}) and (\ref{Eq:Friedmann_2}) as follows:
\begin{equation}\label{Eq:RW6}
    3H^{2} = \rho_{m} + \rho_{DE},
\end{equation}
\begin{equation}\label{Eq:RW7}
    2\dot{H} = - \left(\rho_{m} + p_{m} + \rho_{DE} + p_{DE} \right),
\end{equation}
where we represent  the energy density and pressure of DE $\rho_{DE}$ and $p_{DE}$ as
\begin{equation}\label{Eq:RW8}
    \rho_{DE} = \frac{1}{2}\dot{\phi}^{2} + V - 3H^{2}\mathcal{F},
\end{equation}
\begin{equation}\label{Eq:RW9}
    p_{DE} = \frac{1}{2}\dot{\phi}^{2} - V + 2\mathcal{F}'H\dot{\phi} + 2\dot{H}\mathcal{F} + 3H^{2}\mathcal{F},
\end{equation}
and the corresponding EoS parameter $\omega_{DE}$ for DE can be expressed as 
\begin{equation}
    \omega_{DE} \equiv \frac{p_{DE}}{\rho_{DE}}= - 1 + \frac{\dot{\phi}^{2} + 2\mathcal{F}'H\dot{\phi} + 2\dot{H}\mathcal{F}}{\frac{1}{2}\dot{\phi}^{2} + V - 3H^{2}\mathcal{F}}.
\end{equation}
The equations (\ref{Eq:Friedmann_1}) and (\ref{Eq:Friedmann_2}), lead to the continuity equation which is given by
\begin{equation}\label{Eq:con1}
    \dot{\phi}(\ddot{\phi} + 3H\dot{\phi} + V' + 3\mathcal{F}'H^{2}) + \dot{\rho}_{m} + 3H\rho_{m} = 0.
\end{equation}
The above equations can be decoupled as
\begin{equation}\label{Eq:con2a}
    \dot{\rho}_{m} + 3H\rho_{m} = 0,
\end{equation}
\begin{equation}\label{Eq:con3a}
   \ddot{\phi} + 3H\dot{\phi} + V' + 3\mathcal{F}'H^{2} = 0,
\end{equation}
for $t<t_{i}$, where $t_{i}$ is the time when interaction sets in. However at time $t > t_{i}$ we assume an interaction between DM and DE, yielding the continuity equations
\begin{equation}\label{Eq:con2}
    \dot{\rho}_{m} + 3H\rho_{m} = \mathcal{Q},
\end{equation}
\begin{equation}\label{Eq:con3}
   \ddot{\phi} + 3H\dot{\phi} + V' + 3\mathcal{F}'H^{2} = - \frac{\mathcal{Q}}{\dot{\phi}},
\end{equation}
where $\mathcal{Q}$ is the interaction rate and the energy that flow from DM to DE for positive rate ($\mathcal{Q} > 0$) and DE to DM for the negative rate ($\mathcal{Q} < 0$). Different functional forms of interactions are considered to probe the onset era of interaction. There are no strict rules to assume a particular form of interaction, but some phenomenological choices can be made to begin with which are thereafter verified for their suitability in the context of astronomical observations. To name a few  forms of interactions $\mathcal{Q}$ such as (i) $\mathcal{Q} \propto \rho_{m}$ \cite{amendola2007consequences,V_liviita_2008}, (ii)  $\mathcal{Q} \propto \rho_{DE}$ \cite{PhysRevD.85.043007,Yang_2014}, (iii) $\mathcal{Q} \propto (\rho_{m} + \rho_{DE})$ \cite{V_liviita_2008} etc. 

Introducing the effective energy density $\rho_{eff}$ and pressure $p_{eff}$ as
\begin{equation}\label{Eq:reff}
    \rho_{eff} = \rho_{m} + \frac{1}{2}\dot{\phi}^{2} + V - 3H^{2}\mathcal{F},
\end{equation}
\begin{equation}\label{Eq:peff}
    p_{eff} = p_{m} + \frac{1}{2}\dot{\phi}^{2} - V + 2\mathcal{F}'H\dot{\phi} + 2\dot{H}\mathcal{F} + 3H^{2}\mathcal{F},
\end{equation}
the effective EoS parameter $\omega_{eff}$ yields
\begin{equation}
    \omega_{eff} = \frac{p_{eff}}{\rho_{eff}} = - 1 -\frac{2}{3}\frac{\dot{H}}{H^{2}}.
\end{equation}
The EoS parameter $\omega_{eff}$ represents the matter-dominated era for $\omega_{eff} = 0$, whereas radiation-dominated era for $\omega_{eff} = \frac{1}{3}$. It is also helpful to probe a decelerating universe from an accelerating one. For an accelerated universe one requires $\omega_{eff} < - \frac{1}{3}$. Thus the EoS parameter classifies three possible states for the accelerating universe which are (i) quintessence $(-1 < \omega_{eff} < - \frac{1}{3})$, (ii) phantom ($\omega_{eff} < -1$) and (iii) the cosmological constant ($\omega = -1$). Finally, from the first Friedmann equation (\ref{Eq:Friedmann_1}), we get
\begin{equation}
    \Omega_{m} + \Omega_{DE} = 1,
\end{equation}
where $\Omega_{m} = \frac{\rho_{m}}{3H^{2}}$ and $\Omega_{DE} = \frac{\frac{1}{2}\dot{\phi}^{2} + V - 3H^{2}\mathcal{F}}{3H^{2}}$ are the energy density parameters.

\section{Dynamical system for the model}
\label{sec:dyn}
The field equations are highly nonlinear, we explore cosmological solutions at different phases making use of a dynamical system. It is a mathematical framework used to describe how a system evolves over time and the primary goal is to determine the criteria for the stability of the solutions with the critical (or fixed or equilibrium) points. Typically, a standard form of dynamical system is expressed as follows:
\begin{equation}\label{Eq:ds1}
    \dot{\mathbf{x}} = f(\mathbf{x})
\end{equation}
where $\mathbf{x} = (x_{1}, x_{2},.....,x_{n}) \in X$ to be an element of the state space $X \subseteq \mathbb{R}^{n}$ and the function $f : X \rightarrow X$. The overhead dot represents the derivative with respect to some parameter $t\in\mathbb{R}$. The function $f(\mathbf{x})$ is a vector space on $\mathbb{R}^{n}$ such that
\begin{equation}\label{Eq:ds2}
    f(\mathbf{x}) = (f_{1}(\mathbf{x}),f_{2}(\mathbf{x}),....,f_{n}(\mathbf{x})).
\end{equation} 
In general, the above dynamical system (\ref{Eq:ds1}) can be viewed as a system for $n$ variables with $n$ differential equations. The equation $\dot{\mathbf{x}} = f(\mathbf{x})$ is an ordinary differential equation and represents an autonomous system. The autonomous equation (\ref{Eq:ds1}) is said to have a critical point at $\mathbf{x} = \mathbf{x}_0$ if and only if $f(\mathbf{x}_0) = 0$. This implies that the system can remain in that state indefinitely. However, an important question arises: Can the system actually attains such a state, and if so, is the state stable under small perturbations? These considerations naturally lead to the stability analysis of a critical point. The stability of $\mathbf{x}_{0}$ is classified as follows:

\begin{itemize}
    \item[$\bullet$]  {\bf Stable Fixed Point:} The fixed point $\mathbf{x}_{0}$ is stable if, for every $\epsilon > 0$, there exist a $\sigma > 0$ such that for any solution $\psi(t)$ of (\ref{Eq:ds1}) satisfying $||\psi(t_{0}) - \mathbf{x}_0|| < \sigma $, the solution $\psi(t)$ exists for all $t \geq t_{0}$ and it will satisfy $||\psi(t) - \mathbf{x}_{0}|| < \epsilon$ for all $t \geq t_{0}$.

    \item[$\bullet$] {\bf Asymptotically Stable Fixed Point:} A stable fixed point $\mathbf{x}_{0}$ is said to be asymptotically stable if there exists  $\sigma> 0$ such that for any solution  $\psi(t)$ satisfying $||\psi(t_{0}) - \mathbf{x}_{0}|| < \sigma$, then lim$_{t\rightarrow \infty}\psi(t) = \mathbf{x}_{0}$.
\end{itemize}
The difference between stable and asymptotic stable fixed points is that in an asymptotic fixed points all trajectories approach the point while in a stable fixed point all trajectories make a circle near that point. If a fixed point $\mathbf{x}_{0}$ is neither stable nor asymptotically stable, then it is unstable, meaning that trajectories starting near $\mathbf{x}_{0}$ thereafter diverge away from it.

Having defined a concept of stability, there are several methods which can be used to study the stability properties of critical points. The most common techniques are linear stability theory, Kosambi-Cartan-Chern (KCC) theory, Lyapunov stability, and centre manifold theory. However, we performed the stability analysis using linear stability theory. The basic idea of linear stability theory corresponding to the dynamical system (\ref{Eq:ds1}) is to linearize the system near a fixed point that leads to the Jacobian matrix defined as:
\begin{gather}
   J =  
   \begin{pmatrix}
    \frac{\partial f_{1}}{\partial x_{1}} & \cdots & \frac{\partial f_{1}}{\partial x_{n}}\\
    \vdots & \ddots & \vdots\\
    \frac{\partial f_{n}}{\partial x_{1}} & \cdots & \frac{\partial f_{n}}{\partial x_{n}}
    \end{pmatrix}
\end{gather}

which is called the stability matrix. The eigenvalues of the Jacobian matrix $J$ are evaluated at the critical points $\mathbf{x}_{0}$ which contain information about the stability of the system. If the real parts of all eigenvalues are negative, the fixed point is stable or an attractor. Conversely, if the real parts of all eigenvalues are positive, the fixed point is unstable or a repeller. If some eigenvalues have negative real parts while others have positive real parts, the fixed point is classified as a saddle point.

Therefore, we are concerned with possible dynamical system applications for understanding cosmic evolution by rewriting the cosmological equations (\ref{Eq:Friedmann_1}) and (\ref{Eq:Friedmann_2}) into an autonomous system of equations. For this, we first rewrite the Friedmann equation (\ref{Eq:Friedmann_1}) in dimensionless form which is given by
\begin{equation}
    1 = \frac{\rho_{m}}{3H^{2}} + \frac{\dot{\phi}^{2}}{6H^{2}} + \frac{V(\phi)}{3H^{2}} - \mathcal{F}(\phi).
\end{equation}
To obtain the corresponding dynamical system of the model, we define the following useful dimensionless variables: 
\begin{equation}\label{Eq:dynvaria}
    x = \frac{\dot{\phi}}{\sqrt{6}H},\hspace{1cm} y = \frac{\sqrt{V}}{\sqrt{3}H}, \hspace{1cm} u = - \mathcal{F}.
\end{equation}
Using the above set of dynamical variables, the Friedmann equation (\ref{Eq:Friedmann_1}) can be expressed as
\begin{equation}\label{Eq:Omega_m}
    \Omega_{m} =  1 - x^{2} - y^{2} - u, 
\end{equation}
where $\Omega_{m} \equiv \frac{\rho_{m}}{3H^{2}}$ is the energy density parameter of matter. In equation (\ref{Eq:Omega_m}), $x^{2}$ is the kinetic energy density of the scalar field $\phi$ and $y^{2} + u$ is the effective potential energy density. The total energy density of the scalar field is given by
\begin{equation}\label{Eq:Omega_phi}
    \Omega_{DE} = x^{2} + y^{2} + u.
\end{equation}

The derivatives of the above dynamical variables with respect to e-folds $N = \ln a$ are derived and then using  $\ddot{\phi}$ from equation (\ref{Eq:con3}), we obtain the dynamical system given by 
\begin{equation}\label{Eq:dyn1}
    \frac{dx}{dN} = - 3x + \frac{\sqrt{6}}{2}\beta y^{2} - \frac{\sqrt{6}}{2}\alpha u - x\frac{\dot{H}}{H^{2}} - \frac{\mathcal{Q}}{\sqrt{6}\dot{\phi}H^{2}},
\end{equation}
\begin{equation}\label{Eq:dyn2}
    \frac{dy}{dN} = - \frac{\sqrt{6}}{2}\beta \; xy - y\frac{\dot{H}}{H^{2}},
\end{equation}
\begin{equation}\label{Eq:dyn3}        
        \frac{du}{dN} = - \sqrt{6}\alpha \; xu,
\end{equation}
where we define 
\begin{equation}
    \beta = -\frac{V'}{V},\hspace{0.5cm} \alpha = -\frac{\mathcal{F}'}{\mathcal{F}}. 
\end{equation}
and furthermore, from the equation (\ref{Eq:Friedmann_2}) for  $p_{m} = 0$, we obtain
\begin{equation}
    \frac{\dot{H}}{H^{2}} = - \frac{3}{2(1 - u)}\left(x^{2} - y^{2} + 1 - u + \frac{2\sqrt{6}}{3}\alpha ux\right).
\end{equation}
It is evident that the dynamical equations (\ref{Eq:dyn1})-(\ref{Eq:dyn3}) are not an autonomous system since $\alpha$ and $\beta$ depend on the scalar field $\phi$. In order to make the system autonomous for a general potential, we can express $\alpha$ and $\beta$ as another dynamical variable and look for an evolution equation governing its dynamics. We derive the dynamical equation for the variables $\alpha$ and $\beta$ which are given by
\begin{equation}\label{Eq:dyn4}        
    \frac{d\beta}{dN} = - \sqrt{6}(\Gamma - 1)\beta^{2}x,
\end{equation}
\begin{equation}\label{Eq:dyn5}        
    \frac{d\alpha}{dN} = - \sqrt{6}(\Theta - 1)\alpha^{2}x,
\end{equation}
where we have defined
\begin{equation}\label{Eq:dyn1a}
    \Gamma = \frac{VV''}{V'^{2}}, \hspace{0.5cm} \Theta = \frac{\mathcal{F}\mathcal{F}''}{\mathcal{F}'^{2}}. 
\end{equation}
The dynamical equations (\ref{Eq:dyn1})-(\ref{Eq:dyn3}) and (\ref{Eq:dyn4})-(\ref{Eq:dyn5}) so obtained are not yet a complete autonomous system unless the parameters $\Gamma$ and $\Theta$ are explicitly determined. To analyze the stability of the system, we consider specific functional forms for the potential $V(\phi)$ and the nonminimal coupling function $\mathcal{F}(\phi)$ which are given by
\begin{equation}\label{Eq:exam}
    V(\phi) = V_{0}e^{-\lambda\phi},\hspace{0.5cm} \mathcal{F}(\phi) = \mathcal{F}_{0}e^{-\xi\phi}
\end{equation}
where $\lambda$ and $\xi$ are arbitrary constants. This scalar potential can give rise to an accelerated expansion, and at the same time, it allows to obtain scaling solutions \cite{PhysRevD.57.4686}. On the other hand, the coupling function $\mathcal{F}(\phi)$ is the most natural and simple choice compatible with the exponential scalar potential.

For a scalar field with an exponential potential and an exponential form of the nonminimal coupling function $\mathcal{F}(\phi)$, as given in equation (\ref{Eq:exam}), the parameters take the specific values $\Gamma = 1$ and $\Theta = 1$. Hence, $\alpha$ and $\beta$ serve merely as parameters rather than dynamical variables. As a result, the system of equations (\ref{Eq:dyn1})-(\ref{Eq:dyn3}) transforms into a fully autonomous system, making it sufficient for conducting a comprehensive dynamical analysis. This allows for a systematic investigation of the stability properties and phase-space structure of the system.

Finally, we express the effective EoS parameter and the deceleration parameter in terms of the dynamical variables as 

\begin{equation}\label{}
     \omega_{eff} = - 1 - \frac{2}{3}\frac{\dot{H}}{H^{2}} = -1 + \frac{1}{(1 - u)}\left(x^{2} - y^{2} + 1 - u + \frac{2\sqrt{6}}{3}\alpha ux\right),
\end{equation}
\begin{equation}\label{}
     q = - 1 - \frac{\dot{H}}{H^{2}} = -1 + \frac{3}{2(1 - u)}\left(x^{2} - y^{2} + 1 - u + \frac{2\sqrt{6}}{3}\alpha ux\right).
\end{equation}

\section{ Critical points and the numerical analysis}
\label{sec:4}
To determine the dynamical features of the corresponding autonomous system, we first determine the critical points or fixed points $(x_{c}, y_{c}, u_{c})$ by imposing the following conditions
\begin{equation}
    \frac{dx}{dN} = 0,\;\;\; \frac{dy}{dN} = 0,\;\;\; \frac{du}{dN} = 0,
\end{equation}
and test the stability of the system at those fixed points through the linear stability theory. We discuss two cases (i) non-interacting case, (ii) interacting case.
\subsection{Non-interaction case $\mathcal{Q} = 0$}
Using $\mathcal{Q} = 0$ in equation (\ref{Eq:dyn1}) we get the following autonomous system:
\begin{equation}\label{Eq:auto1}
    \frac{dx}{dN} = - 3x + \frac{\sqrt{6}}{2}\beta y^{2} - \frac{\sqrt{6}}{2}\alpha u - x\frac{\dot{H}}{H^{2}},
\end{equation}
\begin{equation}\label{Eq:auto1a}
        \frac{dy}{dN} = - \frac{\sqrt{6}}{2}\beta xy - y\frac{\dot{H}}{H^{2}},        
\end{equation}
\begin{equation}\label{Eq:auto1b}        
        \frac{du}{dN} = - \sqrt{6}\alpha xu.
\end{equation}
For the above system (equations (\ref{Eq:auto1})-(\ref{Eq:auto1b})), the critical points are presented in Table \ref{tab:critical_1}. The expressions for the cosmological parameters and the cosmological solutions for each critical point are tabulated in Table \ref{tab:critical_1a} and Table \ref{tab:critical_1b}, respectively.

\begin{table}[h]
    \centering
    \caption{Critical points for the autonomous system of non-interaction case (equations \ref{Eq:auto1}-\ref{Eq:auto1b})}
    \setlength{\tabcolsep}{23pt}
    \renewcommand{\arraystretch}{1.5}
    \begin{tabular}{lcccc}
        \hline
        Critical points & $x_{c}$ & $y_{c}$ & $u_{c}$ & Existence\\
        \hline
        $C_1$ & $-1$ & $0$ & $0$ & Always\\
        $C_2$ & $0$ & $0$ & $0$ & Always\\
        $C_3$ & $1$ & $0$ & $0$ & Always\\
        $C_4$ & $\frac{\beta}{\sqrt{6}}$ & $\sqrt{1-\frac{\beta^{2}}{6}}$ & $0$ & $\beta^{2} \leq 6$\\
        $C_5$ & $\frac{\beta}{\sqrt{6}}$ & $-\sqrt{1-\frac{\beta^{2}}{6}}$ & $0$ & $\beta^{2} \leq 6$\\
        $C_6$ & $\frac{\sqrt{\frac{3}{2}}}{\beta}$ & $-\frac{\sqrt{\frac{3}{2}}}{\beta}$ & $0$ & $|\beta| > 0$\\
        $C_7$ & $\frac{\sqrt{\frac{3}{2}}}{\beta}$ & $\frac{\sqrt{\frac{3}{2}}}{\beta}$ & $0$ & $|\beta| > 0$\\
        \hline
    \end{tabular}    
    \label{tab:critical_1}
\end{table}

\begin{table}
    \centering
    \caption{The physical parameter corresponding to the critical points given in Table \ref{tab:critical_1}}
    \setlength{\tabcolsep}{20pt}
    \renewcommand{\arraystretch}{2}
    \begin{tabular}{lcccc}    
        \hline
        Critical points & $\Omega_{m}$ & $\Omega_{DE}$ & $\omega_{eff}$ & $q$\\
        \hline
        $C_{1},C_{3}$ & $0$ & $1$ & $1$ & $2$\\
        $C_{2}$ & $1$ & $0$  & $0$ & $\frac{1}{2}$\\
        $C_{4},C_{5}$ & $0$ & $1$ & $- 1 + \frac{1}{3}\beta^{2}$ & $- 1 + \frac{1}{2}\beta^{2}$\\
        $C_{6},C_{7}$ & $1 - 3\beta^{2}$ & $3\beta^{2}$  & $0$ & $\frac{1}{2}$\\
        \hline
    \end{tabular}
    \label{tab:critical_1a}
\end{table}

\begin{table*}[!]
    \centering
    \caption{Cosmological solutions of the critical points given in Table \ref{tab:critical_1}}
    \setlength{\tabcolsep}{6pt}
    \renewcommand{\arraystretch}{1.5}
    \begin{tabular}{lccc}
        \hline
        Critical points & Acceleration equation & Scale factor  & Universe phase \\
        \hline
        $C_1,C_3$ & $\dot{H} = - 3H^{2}$ & $a(t) = t_{0}(3t + c_{2})^{\frac{1}{3}}$ & Stiff matter \\
        $C_2$ & $\dot{H} = - \frac{3}{2}H^{2}$ & $a(t) = t_{0}(\frac{3}{2}t + c_{2})^{\frac{2}{3}}$ & Matter-dominated \\
        $C_4,C_5$ & $\dot{H} = - \frac{1}{2}\beta^{2}H^{2}$ & $a(t) = t_{0}(\frac{\beta^{2}}{2}t + c_{2})^{\frac{2}{\beta^{2}}}$ & Power-law expansion\\
        $C_6,C_7$ & $\dot{H} = - \frac{3}{2}H^{2}$ & $a(t) = t_{0}(\frac{3}{2}t + c_{2})^{\frac{2}{3}}$ & Matter scaling \\
        \hline
    \end{tabular}
    \label{tab:critical_1b}
\end{table*}

\begin{figure}[h]
    \centering    \includegraphics[width=0.45\textwidth,height=5cm]{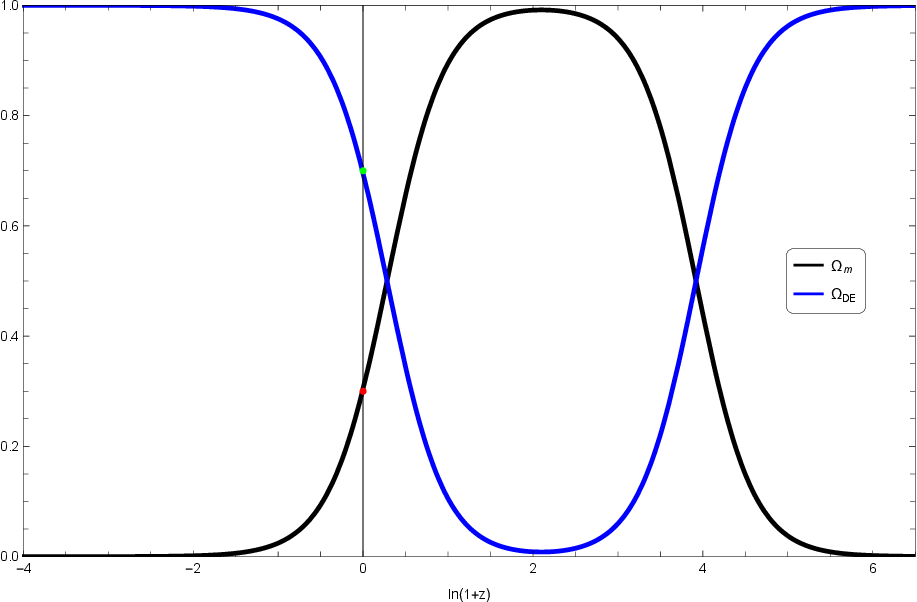}
    \qquad
    \includegraphics[width=0.45\textwidth,height=5cm]{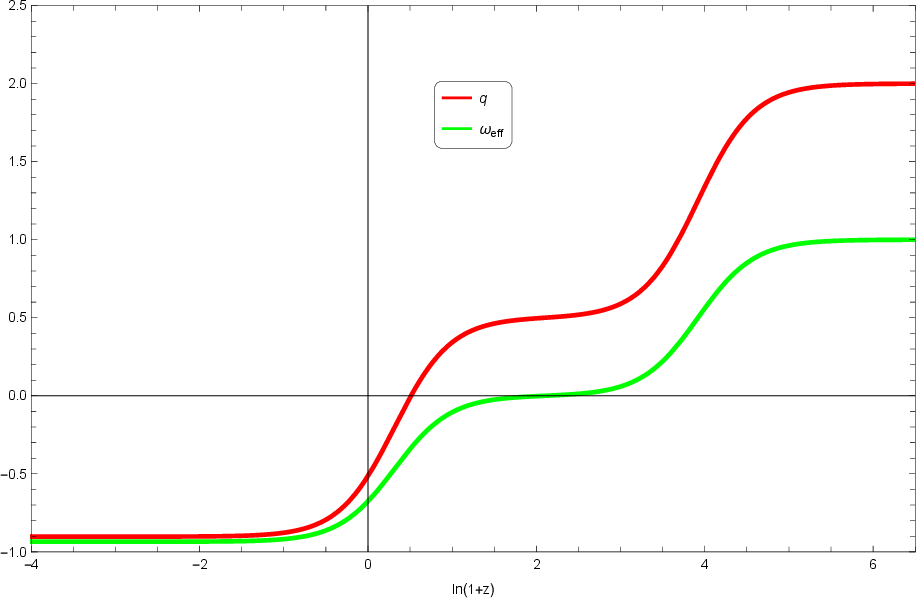}
    \caption{(Left panel) The evolution of the density parameters $\Omega_{m}$(Black curve) and $\Omega_{m}$(Blue curve) for $\alpha = 0.25$ and $\beta = 0.44$ in the non-interacting case. (Right Panel) The variation of the deceleration parameter $q$(Red curve) and effective EoS parameter $\omega_{eff}$(Green curvr) for $\alpha = 0.25$ and $\beta = 0.44$ in the non-interacting case.}
    \label{cos_dia:1}
\end{figure}

\begin{figure}
    \centering    \includegraphics[width=0.4\textwidth,height=5cm]{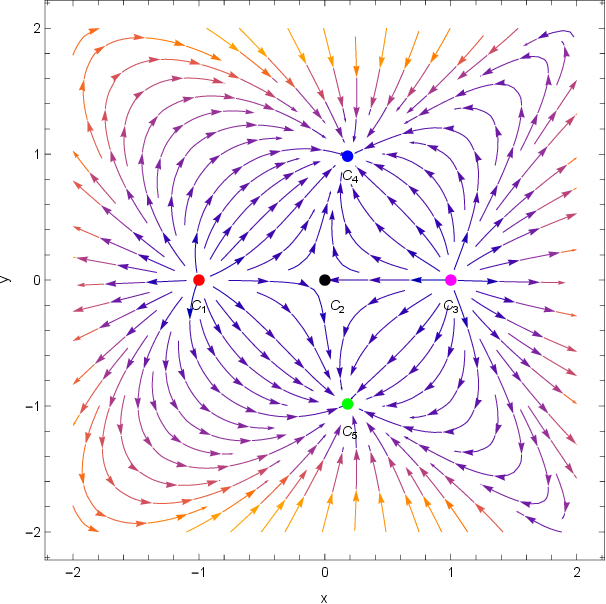}
    \qquad
    \includegraphics[width=0.4\textwidth,height=5cm]{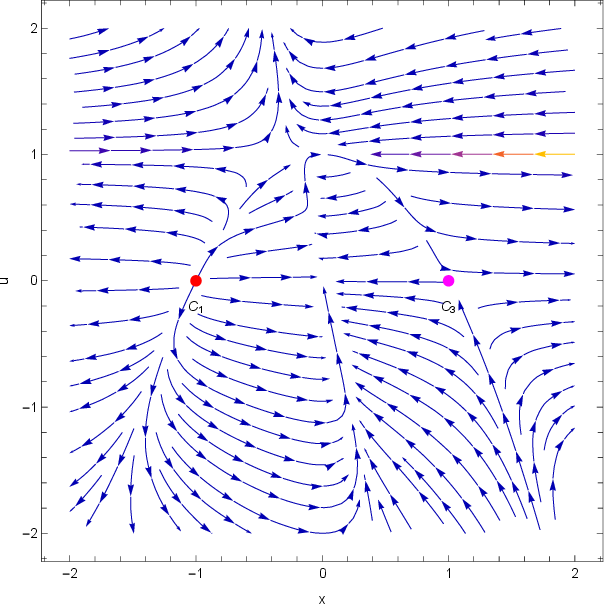}
    \caption{The 2D phase diagram in (left panel) $x-y$ and (right panel) $x-u$ plane for non-interacting case.}
    \label{phase_dia:1}
\end{figure}

We note seven critical points, which are examined to test the nature of the critical points and their stability.
\begin{itemize}
    \item[$\bullet$] {\bf Critical point $C_{1}$}

    At this point, the universe is completely dominated by kinetic energy of the scalar field and the corresponding energy density parameter of matter and scalar field are $\Omega_{m} = 0$ and $\Omega_{DE} = 1$, respectively. Since $\Omega_{m} = 0$, there is no contribution from matter, and the entire energy content of the universe arises from the scalar field. Therefore, this point represents a phase where the universe is completely scalar field-dominated. The deceleration parameter at this point is $q = 2$, which indicates a decelerating expansion and the effective EoS parameter tends to $\omega_{eff} = 1$, which is characteristic of a universe dominated by a stiff fluid. A stiff-fluid EoS corresponds to a scenario where the pressure is equal to the energy density ($p = \rho$), leading to a rapid decelerating expansion of the universe. Such a phase is not viable for late-time cosmic evolution as it does not admit an accelerated expansion. However, stiff-fluid solutions are often relevant in the very early universe, particularly in models where the scalar field plays a significant role in the pre-inflationary or early kinetic-dominated regimes. Due to this, these solutions are typically ignored in discussions of DE and late-time acceleration. The scale factor evolution at this point follows $a(t) = t_{0}(3t + c_{2})^{\frac{1}{3}}$. The power-law expansion confirms that the universe undergone a decelerated evolution assisted by the stiff-fluid-dominated scenario. The eigenvalues at the  point corresponding to the Jacobian metrix of perturbation are 
    \[\left\{\lambda_{1} = 3,\hspace{0.25cm} \lambda_{2} = \sqrt{6}\alpha,\hspace{0.25cm} \lambda_{3} = \frac{1}{2}(6 + \sqrt{6}\beta)\right\}.\]
    The presence of a positive eigenvalues for the critical point predict an unstable mode. Depending on the values of $\alpha$ and $\beta$, the presence of both positive and negative eigenvalues indicate that the critical point may exhibit saddle-like behavior rather than acting as a fully repelling unstable node. In such cases, the system will be unstable along certain directions in phase diagram while allowing transient stability along with other directions, leading to a mixed dynamical behavior characteristic of a saddle point.

    \item[$\bullet$] {\bf Critical point $C_{2}$}
    
    The density parameters for the point are, $\Omega_{m} = 1$ and $\Omega_{DE} = 0$. This solution permits  a matter-dominated phase of the universe, where the entire energy content is attributed to pressureless matter. Such a phase is crucial in cosmology as it corresponds to the period where structure formation takes place. The corresponding deceleration parameter is $q = \frac{1}{2}$ and effective EoS parameter is $\omega_{eff} = 0$. This behavior of the critical point leads to the decelerating phase of the universe and $\omega_{eff}$ that  matches the EoS parameter of non-relativistic matter (dust), confirming that the dynamics at this point are consistent with a matter-dominated era. At this point, the universe expands according to the power-law solution $a(t) = t_{0}(\frac{3}{2}t + c_{2})^{\frac{2}{3}}$. The eigenvalues of the Jacobian matrix for this critical point are
    \[\left\{\lambda_{1} = - \frac{3}{2},\hspace{0.25cm} \lambda_{2} = \frac{3}{2},\hspace{0.25cm} \lambda_{3} = 0\right\}.\]
    The presence of both positive and negative eigenvalues indicates that this critical point behaves as an unstable saddle point. This means that phase-space trajectories pass through this point and they are eventually drawn toward a stable attractor (such as a late-time DE dominated solution). Physically, this reflects the fact that while the universe undergoes a matter-dominated phase, it eventually transits into another phase of expansion, such as a scalar-field-dominated accelerated expansion, making it an intermediate but necessary stage in cosmic evolution. 
    
    \item[$\bullet$] {\bf Critical point $C_{3}$}
    
    For the critical point $C_{3}$, the corresponding density parameters are $\Omega_{m} = 0$ and $\Omega_{DE} = 1$. The  scalar-field-dominated phase obtained from $C_{3}$ permits a decelerating universe where $q = 2$ and the effective EoS parameter $\omega_{eff} = 1$. The eigenvalues for the critical point are
    \[\left\{\lambda_{1} = 3,\hspace{0.25cm} \lambda_{2} = - \sqrt{6}\alpha,\hspace{0.25cm} \lambda_{3} = \frac{1}{2}(6 - \sqrt{6}\beta)\right\}.\]
    For $\alpha > 0$ with an arbitrary $\beta$, this critical point behaves as a saddle. If $\alpha \leq 0$ and $\beta \leq \sqrt{6}$, then this point represents an unstable node in the phase diagram.

    \item[$\bullet$] {\bf Critical point $C_{4}$, $C_{5}$}
    
    At these points, the density parameters are $\Omega_{m} = 0$ and $\Omega_{DE} = 1$ which again represents the scalar-field-dominated phase of the universe. A significant interesting feature of   cosmology emerged as it admits  late-time acceleration of the universe driven by a dynamical DE component. The value of the deceleration parameter at these points is given by $q = - 1 + \frac{1}{2}\beta^{2}$, leading to  cosmic expansion that permits accelerating universe. 
    However, an accelerating universe ($q < 0$) is emerged when $|\beta| < \sqrt{2}$, and the effective EoS parameter $\omega_{eff} = - 1 + \frac{1}{3}\beta^{2}$. For $|\beta| < \sqrt{2}$, the effective EoS satisfies $\omega_{eff} < - \frac{1}{3}$, confirming that the critical  points admit accelerating expansion of the universe. The scenario is consistent with the observed late-time cosmic acceleration which one obtains with the DE. The eigenvalues of these points are
    \[\left\{\lambda_{1} = - \alpha\beta,\hspace{0.25cm} \lambda_{2} = \frac{1}{2}(- 6 +\beta^{2}),\hspace{0.25cm} \lambda_{3} = -3 + \beta^{2}\right\}.\]
    Thus, the critical points exhibit a stable, accelerating expansion for the parameter range: $|\beta| < \sqrt{2}$ and $\alpha\beta > 0$. When $|\beta|$ is sufficiently small, the system behaves like a cosmological constant-dominated universe ($\omega_{eff} \approx - 1$), making it a cosmological model similar to the evolution of the universe with DE.

    \item[$\bullet$] {\bf Critical points $C_{6}$, $C_{7}$}
    
    At the critical points, the density parameters are given by $\Omega_{m} = 1 - 3\beta^{2}$ and $\Omega_{DE} = 3\beta^{2}$. A physically interesting cosmological model is permitted with the critical points when $\beta$  satisfy an inequality $|\beta| \leq \sqrt{\frac{1}{3}}$. This ensures that the matter density parameter $\Omega_{m}$ remains non-negative. The corresponding deceleration parameter is $q = \frac{1}{2}$, which indicates a decelerating universe.  Additionally, the effective EoS parameter is $\omega_{eff} = 0$. Since this effective EoS parameter matches the EoS of pressureless matter ($\omega_{m} = 0$), these solutions admit a universe with  matter scaling solution. In such a scenario, the energy density of the scalar field $\phi$ scales proportionally with the matter density, ensuring that the relative energy contribution of the scalar field does not dominate over time. The eigenvalues of the Jacobian matrix for these critical points are
    \[\Big\{\lambda_{1} = - 3\frac{\alpha}{\beta}, \lambda_{2} = - \frac{3}{4}\left(1 + \sqrt{\frac{24}{\beta^{2}}-7}\right), \lambda_{3} = - \frac{3}{4}\left(1 - \sqrt{\frac{24}{\beta^{2}}-7}\right)\Big\}.\]
    When $|\beta| \leq \sqrt{\frac{1}{3}}$ the eigenvalue $\lambda_{3} > 0$. Since at least one eigenvalue is positive, the critical points are classified as unstable nodes. This implies that any small perturbation around the points will grow over time, preventing the system from settling into a stable equilibrium. In a cosmological context, this indicates that the matter scaling solutions do not serve as long-term attractors but it can act as transient states during the evolution of the universe.
\end{itemize}    
Therefore, we note unstable solutions at the critical points $C_{1}$ and $C_{3}$ dominated by stiff-fluid like DE. The solutions correspond to a decelerating expansion in the very early universe. In addition, we found a critical point $C_{2}$ that is a saddle point and describes a phase dominated by DM. Furthermore, we have found final attractors at the critical points $C_{4}$ and $C_{5}$ for $|\beta| < \sqrt{2}$ that represent the DE-dominated epoch with cosmic acceleration. Also, the scaling solutions $C_{6}$ and $C_{7}$ which are unstable points for $|\beta| \leq \sqrt{\frac{1}{3}}$ and represent the matter scaling solutions. In order to better understand, we shall numerically integrate the differential equations and find out the evolution of cosmological parameters. The phase diagram is also plotted. In Figure \ref{cos_dia:1}, the evolution plot for cosmological parameters has been given for $\alpha = 0.25$ and $\beta = 0.44$ in the non-interacting case. From the figure, it is evident that the universe is accelerating at the present epoch. The present value of the effective EoS parameter is found to satisfy $- 1 < \omega_{total} < - \frac{1}{3}$. In Figure \ref{phase_dia:1}, the 2D phase portrait is plotted. The portrait shows the trajectory behaviour, first going from the repeller point $C_{1}$ to the saddle point $C_{2}$ and subsequently from $C_{2}$ to the stable point $C_{4}$.

\subsection{Interacting case $\mathcal{Q} \neq 0$}
In this subsection, we study the phase space analysis of the cosmological models when interaction among the fluids sets in. We consider linear interaction $\mathcal{Q} = \eta H\rho_{m}$, where $\eta$ is the strength of the interaction. We get the following autonomous system:
\begin{equation}\label{Eq:auto20}
    \frac{dx}{dN} = - 3x + \frac{\sqrt{6}}{2}\beta y^{2} - \frac{\sqrt{6}}{2}\alpha u - \frac{1}{2x}\eta(1-x^{2}-y^{2}-u) - x\frac{\dot{H}}{H^{2}},
\end{equation}
\begin{equation}\label{Eq:auto21}
    \frac{dy}{dN} = - \frac{\sqrt{6}}{2}\beta xy - y\frac{\dot{H}}{H^{2}},
\end{equation}
\begin{equation}\label{Eq:auto22}
    \frac{du}{dN} = - \sqrt{6}\alpha xu.
\end{equation}
The autonomous system (equations (\ref{Eq:auto20})-(\ref{Eq:auto22})) has a singularity at $x = 0$ and to remove this, one has to multiplied by the positive defined term $x^{2}$. Hence, the above system reduces to
\begin{equation}\label{Eq:auto20a}
    \frac{dx}{dN} = - 3x^{3} + \frac{\sqrt{6}}{2}\beta x^{2} y^{2} - \frac{\sqrt{6}}{2}\alpha x^{2} u - \frac{1}{2}\eta x(1-x^{2}-y^{2}-u) - x^{3}\frac{\dot{H}}{H^{2}},
\end{equation}
\begin{equation}\label{Eq:auto21a}
    \frac{dy}{dN} = - \frac{\sqrt{6}}{2}\beta x^{3}y - x^{2}y\frac{\dot{H}}{H^{2}},
\end{equation}
\begin{equation}\label{Eq:auto22a}
    \frac{du}{dN} = - \sqrt{6}\alpha x^{3}u.    
\end{equation}

The critical points are tabulated in Table \ref{tab:critical_2}, the expressions for the cosmological parameters for each critical point are also shown in Table \ref{tab:critical_2a}.

\begin{table}
    \centering
    \caption{Critical points of the autonomous system (equations  (\ref{Eq:auto20a})-(\ref{Eq:auto22a})) for interacting case}
    \setlength{\tabcolsep}{18pt}
    \renewcommand{\arraystretch}{2.5}
    \begin{tabular}{l c c c c}
        \hline
        Critical points & $x_{c}$ & $y_{c}$ & $u_{c}$ & Existence\\
        \hline
        $F_1$ & $0$ & $1$ & $0$ & Always\\
        $F_2$ & $0$ & $-1$ & $0$ & Always\\
        $F_3$ & $-1$ & $0$ & $0$ & Always\\
        $F_4$ & $1$ & $0$ & $0$ & Always\\
        $F_c$ & $0$ & $y_{c*}$ & $u_{c*}$ & Always\\
        $F_5$ & $\frac{\beta}{\sqrt{6}}$ & $\sqrt{1-\frac{\beta^{2}}{6}}$ & $0$ & $|\beta| \leq \sqrt{6}$\\
        $F_6$ & $\frac{\beta}{\sqrt{6}}$ &  $-\sqrt{1-\frac{\beta^{2}}{6}}$ & $0$ & $|\beta| \leq \sqrt{6}$\\              
        $F_7$ & $\frac{3 - \eta}{\sqrt{6}\beta}$ & $\frac{3 - \eta}{\sqrt{6}\beta}\left(\sqrt{1 + \frac{2\beta^{2}\eta}{(3 - \eta)^{2}}}\right)$ & $0$ & $|\beta| > 0$\\
        $F_8$ & $\frac{3 - \eta}{\sqrt{6}\beta}$ & $- \frac{3 - \eta}{\sqrt{6}\beta}\left(\sqrt{1 + \frac{2\beta^{2}\eta}{(3 - \eta)^{2}}}\right)$ & $0$ & $|\beta| > 0$\\
        \hline
    \end{tabular}
    \label{tab:critical_2}
\end{table}

\begin{table}
    \centering
    \caption{The physical parameter corresponding to the critical points given in Table \ref{tab:critical_2}}
    \setlength{\tabcolsep}{0.5pt}
    \renewcommand{\arraystretch}{2}
    \begin{tabular}{lcccc}    
        \hline
        Critical points & $\Omega_{m}$ & $\Omega_{\phi}$ & $\omega_{eff}$ & $q$\\
        \hline
        $F_{1},F_{2}$ & $0$ & $1$ & $-1$ & $-1$\\         
        $F_{3},F_{4}$ & $0$ & $1$ & $1$ & $2$\\
        $F_{c}$ & $1 - y_{c*}^{2} - u_{c*}$ & $y_{c*}^{2} + u_{c*}$ & $- 1 + \frac{1}{(1-u_{c*})}(1 - y_{c*}^{2} - u_{c*})$ & $- 1 + \frac{3}{2(1-u_{c*})}(1 - y_{c*}^{2} - u_{c*})$\\
        $F_{5},F_{6}$ & $0$ & $1$ & $- 1 + \frac{1}{3}\beta^{2}$ & $- 1 + \frac{1}{2}\beta^{2}$\\
        $F_{7},F_{8}$ & $1 - \frac{\eta^{2} - 6\eta + 9 + \beta^{2}\eta}{3\beta^{2}}$ & $\frac{\eta^{2} - 6\eta + 9 + \beta^{2}\eta}{3\beta^{2}}$ & $-\frac{\eta}{3}$ & $\frac{1}{2} - \frac{\eta}{2}$\\
        \hline
    \end{tabular}
    \label{tab:critical_2a}
\end{table}

\begin{figure}[t]
    \centering    \includegraphics[width=0.45\textwidth,height=5cm]{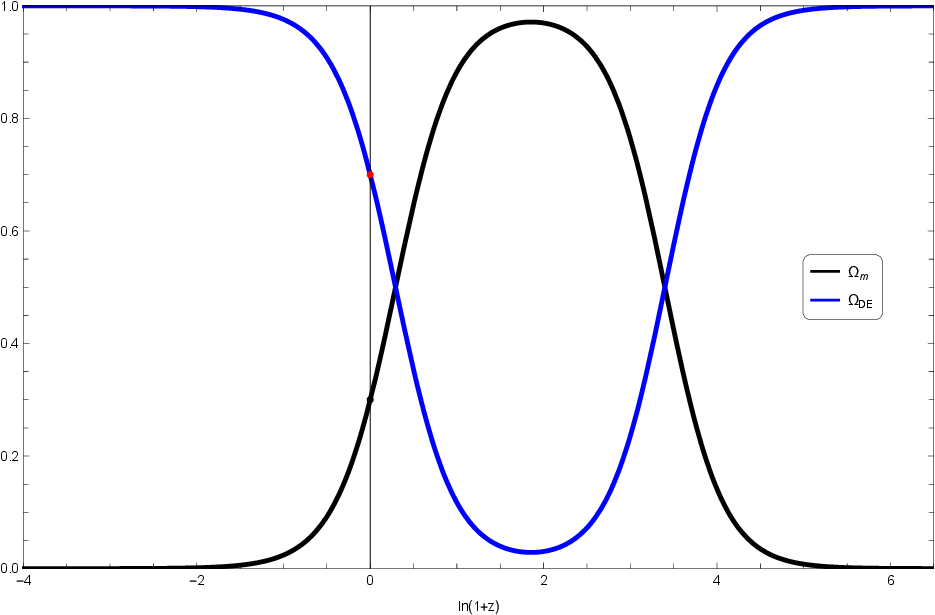}
    \qquad
    \includegraphics[width=0.45\textwidth,height=5cm]{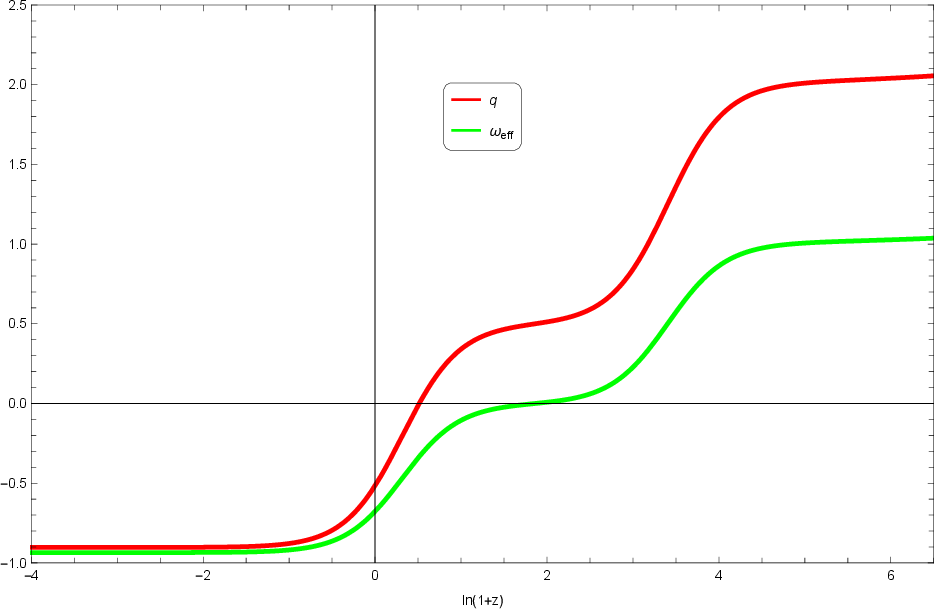}
    \caption{(Left panel) The evolution of the density parameters $\Omega_{m}$(Black curve) and $\Omega_{\phi}$(Blue curve) for interacting case. (Right panel) The variation of the deceleration parameter $q$(Red curve) and effective EoS parameter $\omega_{eff}$(Green curve). The curves corresponding to the values of $\alpha = 0.251$, $\beta = 0.442$ and $\eta = 0.01$.}
    \label{cos_inter_1}
\end{figure}

\begin{figure}
    \centering    
    \includegraphics[width=0.4\textwidth]{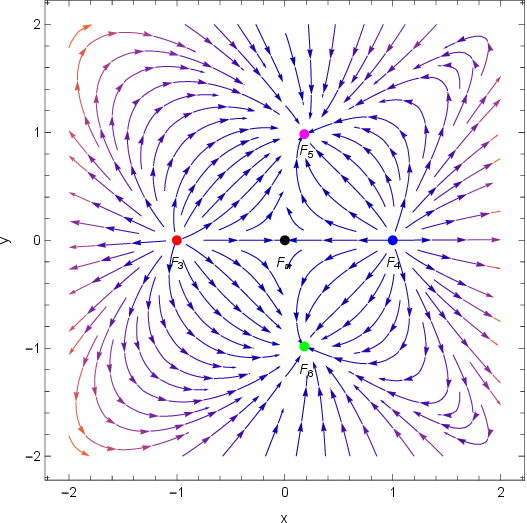}
    \caption{The 2D phase diagram for $\alpha = 0.251$, $\beta = 0.442$ and $\eta = 0.01$ in the presence of interacting.}
    \label{phase_dia:inter1} 
\end{figure}

\begin{figure}
    \centering    \includegraphics[width=0.45\textwidth,height=5cm]{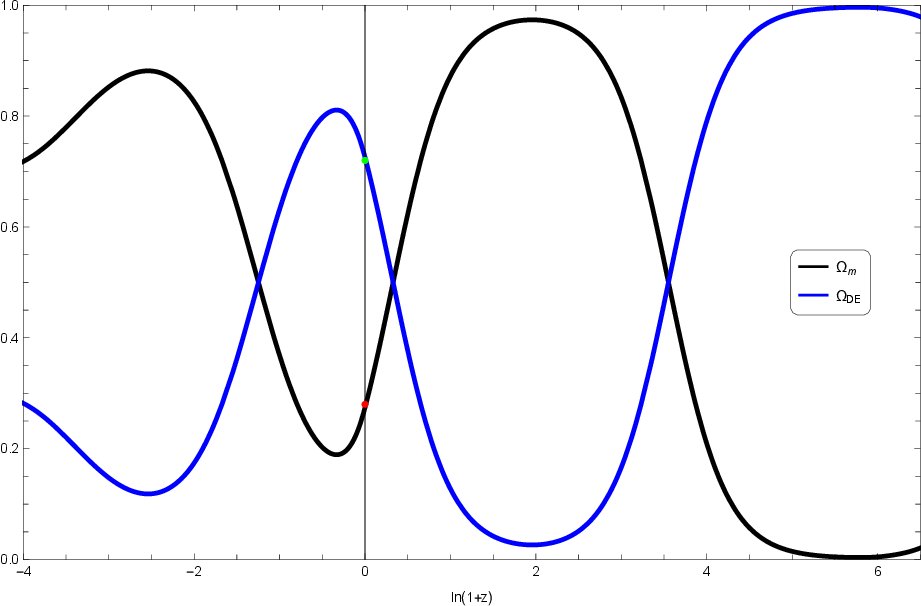}
    \qquad
    \includegraphics[width=0.45\textwidth,height=5cm]{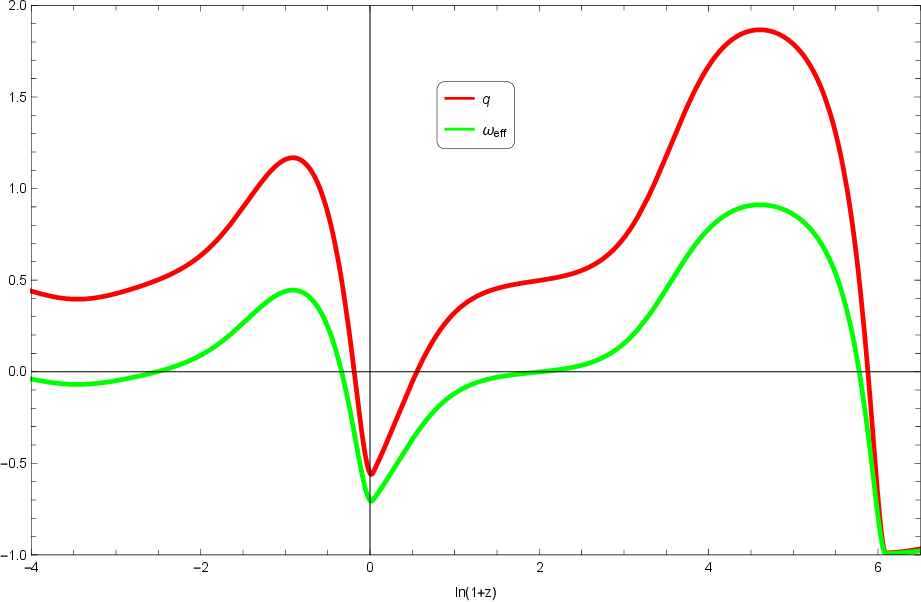}
    \caption{(Left panel) The evolution of the density parameters $\Omega_{m}$(Black curve) and $\Omega_{\phi}$(Blue curve) for interacting case. (Right panel) The variation of the deceleration parameter $q$(Red curve) and effective EoS parameter $\omega_{eff}$(Green curve). The curves corresponding to the values of $\alpha = 0.251$, $\beta = 3.442$ and $\eta = 0.01$.}
    \label{cos_inter_3}
\end{figure}

\begin{figure}
    \centering    \includegraphics[width=0.4\textwidth,height=5cm]{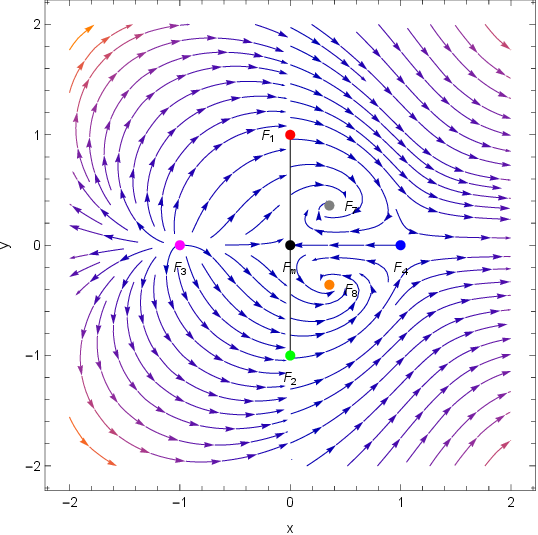}
    \qquad
    \includegraphics[width=0.4\textwidth,height=5cm]{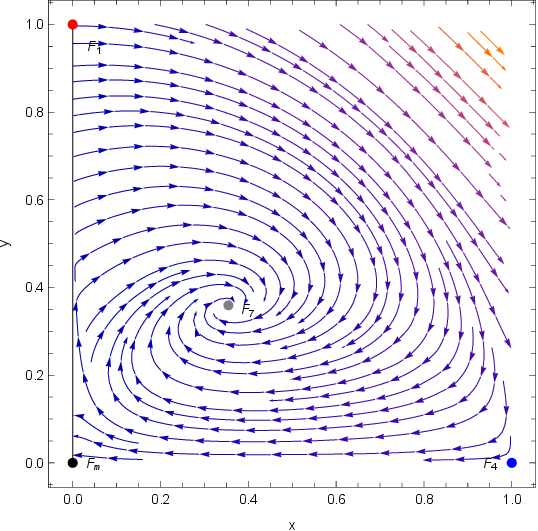}
    \caption{(Left panel) The 2D phase diagram for $\alpha = 0.251$, $\beta = 3.442$ and $\eta = 0.01$ in the presence of interacting. (Right panel) Magnification of subspace of the phase diagram in $x-y$ plane.}
    \label{phase_dia:inter3}
\end{figure}

The stability of the critical points will be studied in this section. 
\begin{itemize}
    \item[$\bullet$] {\bf Critical points $F_{1}$, $F_{2}$}

    These critical points correspond to a phase where the universe is entirely dominated by the potential energy of the scalar field. At these points, the energy density parameters take the values $\Omega_{m} = 0$ and $\Omega_{DE} = 1$. The deceleration parameter and the effective EoS parameter are found to be $q = -1$ and $\omega_{eff} = -1$, respectively, confirming an accelerating expansion. The points $F_{1}$ and $F_{2}$ lead to $\dot{H} = 0$, corresponding to a de Sitter  universe. The eigenvalues of this point vanish and the points are non-hyperbolic in nature. Therefore, the linear stability theory is not suitable to find the nature of the critical points. The stability is shown in the phase diagram  numerically.
    
    \item[$\bullet$] {\bf Critical point $F_{3}$}
    
    This critical point shows a kinetic energy domination of the scalar field and the energy density parameters are $\Omega_{m} = 0$ and $\Omega_{DE} = 1$. Therefore, the universe is a DE-dominated phase at this point. The corresponding deceleration parameter at this point is $q = 2$, indicating a decelerating expansion and the effective EoS parameter is $\omega_{eff} = 1$, confirming that the universe behaves as if dominated by a stiff fluid. The eigenvalues of the Jacobian matrix for this point are
    \[\left\{\lambda_{1} = \sqrt{6}\alpha,\hspace{0.25cm} \lambda_{2} = \frac{1}{2}(6 + \sqrt{6}\beta),\hspace{0.25cm} \lambda_{3} = 3 + \eta\right\}\]
    Since $\eta > 0$, we always have $\lambda_{3} > 0$ and the point cannot be fully stable. Therefore, the point will be either an unstable node or a saddle point in the phase plane depending on the values of $\alpha$ and $\beta$. The condition for an unstable node: $\alpha > 0$ and $\beta > -\sqrt{6}$ and the condition for saddle: (i) $\alpha > 0$ and $\beta < - \sqrt{6}$, (ii) $\alpha < 0$ and $\beta > - \sqrt{6}$.

    \item[$\bullet$] {\bf Critical point $F_{4}$}
    
    Analogous to the point $F_{3}$, the critical point represents a DE-dominated decelerating expansion of the universe with a kinetic energy domination of the scalar field. The corresponding eigenvalues of the Jacobian matrix for this critical point are
    \[\left\{\lambda_{1} = -\sqrt{6}\alpha,\hspace{0.25cm} \lambda_{2} = \frac{1}{2}(6 - \sqrt{6}\beta),\hspace{0.25cm} \lambda_{3} = 3 + \eta\right\}\]
    According to the model $\eta > 0$, we always have $\lambda_{3} > 0$. Therefore, the point cannot be represented as a stable phase of the universe. Thus, the point is either unstable when $\alpha < 0$ and $\beta < \sqrt{6}$ or a saddle point when (i) $\alpha > 0$ and $\beta < \sqrt{6}$, (ii) $\alpha < 0$ and $\beta > \sqrt{6}$.

    \item[$\bullet$] {\bf Surface of critical points $F_{c}$}
    
    On the surface of the parameter space  the critical  points at $F_{c}$ lead to the energy densities which are $\Omega_{m} = 1 - y_{c*}^{2} - u_{c*}$ and $\Omega_{\phi} = y_{c*}^{2} + u_{c*}$. For a physically realistic solutions we get $y_{c*}^{2} \geq 0,\; u_{c*} \geq 0,\; y_{c*}^{2} + u_{c*} \leq 1$. The deceleration parameter $q$ and effective EoS parameter $\omega_{eff}$ are tabulated in Table \ref{tab:critical_2a}. These solutions correspond to accelerated expansion when $3y_{c*}^{2} + u_{c*} > 1$ but decelerating expansion when $3y_{c*}^{2} + u_{c*} < 1$. We note here a matter-dominated decelarating expansion for $y_{c*} \rightarrow 0 \;\; u_{c*} \rightarrow 0$, denoted by $F_{m}$. In the point $F_{m}$, the deceleration parameter is $q = \frac{1}{2}$ and effective EoS parameter $\omega_{eff} = 0$. The eigenvalues of the points on the surface are
    \[\left\{\lambda_{1} = - \frac{1}{2}\eta\left(1 - y_{c*}^{2} - u_{c*}\right),\hspace{0.25cm} \lambda_{2} = \frac{1}{2},\hspace{0.25cm} \lambda_{3} = 0\right\}\]
    Since $\eta > 0$, both positive and negative eigenvalues have been obtained. Hence, the points on the surface permits unstable saddle solution for a physically observable universe.

    \item[$\bullet$] {\bf Critical points $F_{5}$, $F_{6}$}
    
    The critical points exist for $|\beta| \leq \sqrt{6}$ and the corresponding energy density parameters are $\Omega_{m} = 0$ and $\Omega_{DE} = 1$. At these points, the value of the deceleration parameter is given by $q = - 1 + \frac{1}{2}\beta^{2}$. Thus, the universe is accelerating for $|\beta| < \sqrt{2}$ and decelerating for $|\beta| > \sqrt{2}$. If $\beta = 0$, the universe undergoes a perfect exponential expansion with $q = -1$, corresponding to a de Sitter phase. The acceleration decreases as $\beta$ varies as: 
    \[\{\beta \in \mathbb{R}| - \sqrt{2} < \beta < \sqrt{2}, \beta \neq 0\}\]
    The effective EoS parameter in this phase is expressed as $\omega_{eff} = - 1 + \frac{1}{3}\beta^{2}$. If $\beta = 0$, the effective EoS matches that of a cosmological constant, $\omega_{eff} = -1$ and the effective EoS gradually deviates from $-1$ as $\beta$ varies, indicating a departure from the purely vacuum energy-dominated phase. The eigenvalues associated with these points are obtained from the Jacobian matrix of the perturbations
    \[\Big\{\lambda_{1} = - \frac{\alpha\beta^{3}}{6}, \lambda_{2} = \frac{\beta^{2}}{12}(- 6 +\beta^{2}), \lambda_{3} = \frac{\beta^{2}}{6}(-3 + \beta^{2} + \eta)\Big\}\]
    With $\eta > 0$, these points are stable nodes in the phase diagram for the condition $\{\alpha > 0,\; |\beta| < \sqrt{6},\; \beta^{2} +\eta < 3\}$
    and the condition for the unstable point $\{\alpha < 0, \text{or}\; |\beta| > \sqrt{6},\; \text{or}\; \beta^{2} +\eta < 3\}.$    

    \item[$\bullet$] {\bf Critical points $F_{7}$, $F_{8}$}
    
    The critical  points lead to the following:
    \[\Omega_{m} = 1 - \frac{\eta^{2} - 6\eta + 9 + \beta^{2}\eta}{3\beta^{2}},\hspace{0.25cm} \Omega_{\phi} = \frac{\eta^{2} - 6\eta + 9 + \beta^{2}\eta}{3\beta^{2}}.\]
    A physically acceptable cosmological model is obtained at the critical points when $\eta$ satisfy $\eta \in (3-\beta^{2},3)$ and $\beta \neq 0$. The corresponding deceleration parameter $q = \frac{1}{2} - \frac{\eta}{2}$ and effective EoS parameter $\omega_{eff} = - \frac{\eta}{3}$. The eigenvalues of the points are tabulated in Table \ref{tab:eigenvalues}. We obtain numerical solution, and its stability is tested by the phase space analysis.
    \begin{table}
        \centering
        \caption{Eigenvalues of the critical points $F_{7}$, $F_{8}$}
        \renewcommand{\arraystretch}{2.5}
        \begin{tabular}{|c|}
        \hline
        $\lambda_{1} = - \frac{\alpha(3 - \eta)^{3}}{6\beta^{3}},$\\
        $\lambda_{2} = \frac{(3 - \eta)\left(2\beta^{8}\eta + 3\beta^{6}(\eta^{2} - 2\eta - 3) + \sqrt{\beta^{10}\left(- 8(- 3 + \eta)^{5} + 4\beta^{6}\eta^{2} - 4\beta^{4}\eta(\eta^{2}-18\eta + 45) - 3\beta^{2}(3-\eta)^{2}(5\eta^{2} - 38\eta + 21)\right)}\right)}{24\beta^{8}}$\\
        $\lambda_{3} = \frac{(3 - \eta)\left(2\beta^{8}\eta + 3\beta^{6}(\eta^{2} - 2\eta - 3) - \sqrt{\beta^{10}\left(- 8(- 3 + \eta)^{5} + 4\beta^{6}\eta^{2} - 4\beta^{4}\eta(\eta^{2}-18\eta + 45) - 3\beta^{2}(3-\eta)^{2}(5\eta^{2} - 38\eta + 21)\right)}\right)}{24\beta^{8}}$\\
        \hline 
        \end{tabular}    
        \label{tab:eigenvalues}
    \end{table}   
\end{itemize}
 In this case, we get potential energy dominated phase for the points $F_{1}$ and $F_{2}$,  which admits de Sitter expansion. For the points $F_{3}$ and $F_{4}$ we obtain unstable node and saddle points depending on $\alpha$ and $\beta$. The universe is in a decelerating phase dominated by stiff fluid like DE. A decelerating matter-dominated saddle point is found to exist at the critical points $F_{m}$ which lies on the surface $F_{c}$. A stable solutions at the critical points $F_{5}$ and $F_{6}$ are noted when $|\beta| < \sqrt{2}$ accommodating late-time acceleration dominated by DE. Thus, the evolution of the universe gives a scenario where  initially the universe evolves from the repeller point $F_{3}$ then to the saddle point $F_{m}$ and subsequently from $F_{m}$ to the stable critical point either to $F_{5}$ or $F_{6}$. In Figures \ref{cos_inter_1} and \ref{phase_dia:inter1}, we plot basic cosmological parameters and the phase  diagram  for the values of $\alpha = 0.251$, $\beta = 0.442$ and $\eta = 0.01$, respectively. It is also found that a scaling solution at the critical points $F_{7}$ and $F_{8}$ exist. We obtain an interesting  cosmological model which evolves following a chain  $F_{1}\rightarrow F_{4}\rightarrow F_{m}\rightarrow F_{c} \rightarrow F_{7}$ for the values of the parameters $\alpha = 0.251$, $\beta = 3.442$ and $\eta = 0.01$. The corresponding cosmological parameters and phase diagram are plotted in Figures \ref{cos_inter_3} and \ref{phase_dia:inter3}, respectively. From the phase diagram, it is evident that the universe transits from accelerating phase (the critical point $F_{1}$) to a decelerating phase ($F_{4}$) dominated by the scalar field. It is noted  that a matter-dominated universe ($F_{m}$) transits  to an accelerating phase of expansion along the surface $F_{s}$. Finally, the universe transits to a decelarating phase $F_{7}$ in the near future, which supports the recent predictions obtained from Dark Energy Spectroscopic Instrument (DESI) collaboration \cite{desicollaboration2025datarelease1dark,desicollaboration2025desidr2resultsi,desicollaboration2025desidr2resultsii}. The above plots  of evolution are interesting which not only permits the early universe, matter dominated and accelerating phases of the universe. We obtain a new observational feature which is predicted from the analysis that the present accelerating universe because of DE may not expand forever. In future, the universe may flip from accelerating to decelerating phase supporting the recently announced result from DESI mission. The present value of the effective EoS parameter obtained in our model permits  a realistic cosmological scenario for the inequalities $- 1 < \omega_{total} < - \frac{1}{3}$.

\section{Conclusions}
\label{sec:conclusions}
In the paper, we present a dynamical scenario of the universe in a non-minimal scalar non-metricity theory of gravity in a flat FRW geometry. The dynamics of the universe is analyzed here employing dynamical system analysis of the field equations. We rewrite the equations of the dynamical system using dimensionless variables. Each critical point can be connected to the major cosmological epoch of the universe and we test the stability of the solutions analyzing the local asymptotic behavior around small perturbations. The cosmological implications of dynamical system analysis are critically examined with or without interaction between DM and the scalar field, which is non-minimally coupled to the geometry.

In the non-interacting case, the critical points $C_{1}$ and $C_{3}$ are found to dominate by DE similar to stiff-fluid. In this scenario, the universe is decelerating in the very early universe. An unstable solution and a saddle point are found to exist at the critical points for a set of parameters $\alpha$ and $\beta$. At the critical point $C_{2}$ a saddle point is found that describes a phase which is a DM dominated phase. The stable points are found at $C_{4}$ and $C_{5}$ for $|\beta| < \sqrt{2}$, describing late-time cosmic expansion. A matter scaling solution has been found at the critical points $C_{6}$ and $C_{7}$, the unstable nature is found when $|\beta| \leq \frac{1}{3}$. In Figures \ref{cos_dia:1} and \ref{phase_dia:1}, numerical solutions corresponding to the cosmological parameters and the projection of the phase diagram on the $x-y$ and $x-u$ planes are plotted for $\alpha = 0.25$, $\beta = 0.44$, respectively. From the plot, it is found that a matter dominated phase and two DE-dominated phases exist which represent an early and a late expansion of the universe. The variation of the deceleration parameter and the effective EoS parameter admits a transition from deceleration to acceleration phase at the present epoch with  $q = - \frac{1}{2}$ and $\omega_{eff} = -0.73$. The trajectory drawn in the phase diagram shows a transition from the point $C_{1}$ to the point $C_{4}$ via a matter dominated point $C_{2}$.

However, in the case of interaction, the phase diagram Figure \ref{phase_dia:inter1} shows that the universe transits from an unstable point $F_{3}(F_{4})$ called DE dominated phase to a matter dominated phase at $F_{m}$ and subsequently the universe transits to a late-time accelerating phase represented by $F_{5}(F_{6})$ from $F_{m}$. We obtain another interesting result in the interacting picture, for given values of the parameters $\alpha = 0.251$, $\beta = 3.442$ and $\eta = 0.01$. A potential energy dominated unstable point at the critical points $F_{1}$ and $F_{2}$ exists for the scalar field. The above points represent the early accelerating expansion and then to an unstable phase of the universe before the onset of the kinetic energy domination of the scalar field represented by the points $F_{4}$. It is further noted that a transition from matter-dominated universe to a scaling solution exists on the surface $F_{c}$ which represents both accelerating and decelerating phases of the universe, depending on $y_{c}$ and $u_{c}$. Finally, in the near future a decelerating phase is found to exist for the critical points $F_7$ and $F_{8}$ followed by an accelerating present universe, which supports the prediction of DESI. The cosmological parameters are shown in Figure \ref{cos_inter_3}. The trajectory behaviour is shown in the Figure \ref{phase_dia:inter3}.


\acknowledgments

BCR and BCP express their sincere gratitude to the IUCAA Center for Astronomy Research and Development (ICARD), Department of Physics, North Bengal University, for providing invaluable research facilities. BCR gratefully acknowledges the support of the Ministry of Social Justice and Empowerment, Government of India, for the fellowship awarded during the course of this research. BCP extends heartfelt thanks to the Science and Engineering Research Board (SERB), Department of Science and Technology (DST), Government of India, for the project grant (F. No. CRG/2021/000183).



\bibliographystyle{JHEP}
\bibliography{References.bib}





\end{document}